\documentclass[preprint,12pt]{elsarticle}
\usepackage{lineno,subfigure}
\usepackage{color}
\newcommand{\Oh}[1]{\ensuremath{\mathcal{O}\!\left({#1}\right)}}
\newcommand{\depth}{\ensuremath{\mathsf{depth}}}
\newcommand{\nodes}{\ensuremath{\mathsf{nodes}}}
\newcommand{\leaves}{\ensuremath{\mathsf{leaves}}}
\newcommand{\wmm}{\texttt{WMM}}	
	
\newcommand{\tablen}{\texttt{TABLE}}

\newcommand{\B}{\mathcal{B}}
\renewcommand{\log}{\lg}

\newtheorem{theorem}{Theorem}
\newtheorem{corollary}[theorem]{Corollary}

\begin{document}

\begin{frontmatter}

\title{Efficient and Compact Representations of Some Non-Canonical Prefix-Free Codes}

\author[udc]{Antonio Fari\~na}
\author[udp,cebib]{Travis Gagie\corref{cor}}
\author[lodz]{Szymon Grabowski}
\author[upo,cnr]{\mbox{Giovanni Manzini}}
\author[cebib,imfd,chile]{Gonzalo Navarro}
\author[ebay]{Alberto Ord\'o\~nez}

\cortext[cor]{Corresponding author: {\tt travis.gagie@dal.ca}; Faculty of 
Computer Science, Dalhousie University, 6050 University Avenue, PO BOX 15000,
Halifax, Nova Scotia B3H 4R2, Canada.}

\address[udc]{Universidade da Coru\~na and Centro de Investigaci\'on CITIC, A Coru\~na, Spain}
\address[udp]{Dalhousie University, Canada}
\address[cebib]{Center for Biotechnology and Bioengineering (CeBiB), Chile}
\address[lodz]{Institute of Applied Computer Science, Lodz University of Technology, Poland}
\address[upo]{Department of Computer Science, University of Pisa, Italy}
\address[cnr]{IIT-CNR, Pisa, Italy}
\address[imfd]{Millennium Institute for Foundational Research on Data (IMFD), Chile}
\address[chile]{Department of Computer Science, University of Chile, Chile}
\address[ebay]{Pinterest Inc., CA, USA}

\begin{abstract}
For many kinds of prefix-free codes there are efficient and compact alternatives to the traditional tree-based representation.  Since these put the codes into canonical form, however, they can only be used when we can choose the order in which codewords are assigned to symbols.  In this paper we first show how, given a probability distribution over an alphabet of $\sigma$ symbols, we can store an optimal alphabetic prefix-free code in $\Oh{\sigma \log L}$ bits such that we can encode and decode any codeword of length $\ell$ in $\Oh{\min (\ell, \log L)}$ time, where $L$ is the maximum codeword length.  With $\Oh{2^{L^\epsilon}}$ further bits, for any constant $\epsilon>0$, we can encode and decode $\Oh{\log \ell}$ time.  We then show how to store a nearly optimal alphabetic prefix-free code in \(o (\sigma)\) bits such that we can encode and decode in constant time.  We also consider a kind of optimal prefix-free code introduced recently where the codewords' lengths are non-decreasing if arranged in lexicographic order of their reverses.  We reduce their storage space to $\Oh{\sigma \log L}$ while maintaining encoding and decoding times in $\Oh{\ell}$. We also show how, with $\Oh{2^{\epsilon L}}$ further bits, we can encode and decode in constant time.
All of our results hold in the word-RAM model.
\end{abstract}


\end{frontmatter}

%\linenumbers

\section{Introduction}
\label{sec:introduction}

Prefix-free codes are a fundamental tool in data compression;
they are used in one form or another in almost every compression tool. 
Prefix-free codes allow assigning variable-length codewords to symbols 
according to their probabilities in a way that the encoded stream can be
decoded unambiguously \cite[Ch.\ 5]{CT06}. Their best-known representative, 
Huffman codes \cite{Huf52}, yield the optimal encoded file size for a given 
probability distribution. Fast encoding and decoding algorithms for prefix-free 
codes are then of utmost relevance. When the source alphabet is large (e.g., in
word-based natural language compression \cite{Mof89,ZMNBY00}, East Asian or
numeric alphabets) or when the text is short compared to the alphabet (e.g.,
for compression boosting \cite{FGMS05} or adaptive compression \cite{BFNP07}),
a second concern is the space spent in storing the codewords of all the source 
symbols, because it could outweigh the compression savings.

The classical encoding and decoding algorithms for a codeword of length $\ell
\le L$ take in the word-RAM model $\Oh{1}$ and $\Oh{\ell}$ time, respectively, using $\Oh{\sigma L}$
bits of space, where $\sigma$ is the size of the source alphabet and $L$ is
the maximum codeword length. For encoding we just store each codeword in plain 
form, whereas for decoding we use a binary tree $\B$ where each leaf corresponds
to a symbol and the path from the root to the leaf spells out its code, if we 
interpret going left as a $0$ and going right as a $1$. Faster decoding is 
possible if we use the so-called canonical codes, where the leaves are sorted 
left-to-right by depth, and by symbol upon ties \cite{SK64}. Canonical codes 
enable $\Oh{\log L}$-time encoding and decoding while using 
$\Oh{\sigma\log\sigma}$ bits of space, again in the word-RAM model. 
In theory, both encoding and decoding 
can be done even in constant time with canonical codes \cite{GNNO15}.

Both the original and the canonical Huffman codes achieve optimality by 
reordering the leaves as necessary. There are applications for which the 
codes must be so-called alphabetic, that is, the leaves must respect, 
left-to-right, the alphabetic order of the source symbols. This allows
lexicographically comparing strings directly in compressed form, which enables
lexicographic data structures on the compressed strings \cite{BNO12,MPBCCN15}
and compressed data structures that represent point sets as sequences of 
coordinates \cite{Nav14}. Optimal alphabetic (prefix-free) codes achieve 
codeword lengths close to those of Huffman 
codes \cite{HT71}. Interestingly, since the mapping between symbols and leaves 
is fixed, alphabetic codes need only store the topology of the binary tree 
$\B$ used for decoding, which can be represented more succinctly than optimal 
prefix-free codes, in $\Oh{\sigma}$ bits \cite{MR01}, so that encoding and 
decoding can still be done in time $\Oh{\ell}$ \cite{GNNO15}. As far as we 
know, there are no equivalents to the fast and compact representations of 
canonical codes for alphabetic codes.

There are other cases where canonical prefix-free codes cannot be used.
Wavelet matrices, for example, serve as compressed representations of discrete 
grids and sequences over large alphabets \cite{CNO15}. They are compressed with
an optimal prefix-free code where the codewords' lengths are non-decreasing if 
arranged in lexicographic order of their {\em reverses}. They 
represent the code in $\Oh{\sigma L}$ bits, and encode and decode a codeword 
of length $\ell$ in time $\Oh{\ell}$. 

\paragraph{Our contributions}

In Section~\ref{sec:alphabetic} we show how, given a probability distribution,
we can store an optimal alphabetic prefix-free code in $\Oh{\sigma \log L}$ 
bits such that we can encode and decode any codeword of length $\ell$ in 
$\Oh{\min (\ell, \log L)}$ time. This time decreases to $\Oh{\log \ell}$ 
if we use $\Oh{2^{L^\epsilon}}$ additional bits, for any constant $\epsilon>0$.
We then show in Section~\ref{sec:alphabetic2}
how to store a nearly optimal alphabetic prefix-free code in 
\(o (\sigma)\) bits such that we can encode and decode in constant time.  
These, and all of our results, hold in the word-RAM model.

In Section~\ref{sec:matrices} we consider the optimal prefix-free codes used
for wavelet matrices \cite{CNO15}. We show how to store such a code in 
$\Oh{\sigma \log L}$ bits and still encode and decode any symbol in 
$\Oh{\ell}$ time. We also show that, using $\Oh{2^{\epsilon L}}$ further bits, 
we can encode and decode in constant time.
Our first variant is simple enough to be implementable. Our
experiments show that on large alphabets it uses 20--30 times less space 
than a classical implementation, at the price of being 10--20 times slower at 
encoding and 10--30 at decoding.

An early version of this paper appeared in {\em Proc. SPIRE 2016} 
\cite{SPIRE16}. This extended version includes much more detailed 
explanations as well as new results for fast encoding and decoding of
optimal alphabetic codes (Section~\ref{sec:alphabetic}).

\section{Basic Concepts}

\subsection{Assumptions}
\label{sec:assumptions}

Our results hold in the word-RAM model, where the computer word has $w$ bits 
and all the basic arithmetic and logical operations can be carried out in 
constant time. We assume for simplicity that the maximum codeword length is 
$L = \Oh{w}$, so that any codeword can be accessed in $\Oh{1}$ time. We assume 
binary codewords, which are the most popular because they provide the best 
compression, though our results generalize to larger alphabets.

We generally express the space in bits, but when we say $\Oh{x}$ space, we
mean $\Oh{x}$ words of space, that is, $\Oh{xw}$ bits. 

By $\lg$ we denote the logarithm to the base $2$ by default.

\subsection{Basic data structures}
\label{sec:basics}

\paragraph{Predecessors}
This predecessor problem consists in building a data structure on the integers
$0 \le x_1 < x_2 < \cdots < x_n < U$ such that later, given an integer $y$,
we return the largest $i$ such that $x_i \le y$. In the RAM model, with
$\log U = \Oh{w}$, it can be 
solved with structures using $\Oh{n\log U}$ bits in $\Oh{\log\log U}$ time, as 
well as in $\Oh{\log_w n}$ time, among other tradeoffs \cite{PT06}. It is 
also possible to find the answer in time $\Oh{\log i}$ using exponential 
search.

\paragraph{Bitmaps}
A bitmap $B[1..n]$ is an array of $n$ bits that supports two operations:
$rank_b(B,i)$ counts the number of bits $b \in \{0,1\}$ in $B[1..i]$, and 
$select_b(B,j)$ gives the position of the $j$th $b$ in $B$ (we use $b=1$ by
default). Both operations can be supported in constant time if we store $o(n)$ 
bits on top of the $n$ bits used for $B$ itself \cite{Cla96,Mun96}. When $B$ 
has $m$ $1$s and $m \ll n$ or $n-m \ll n$, it can be represented
in compressed form, using $m\log(n/m) + O(m+n/\log^c n)$ bits in total for any 
$c$, so that $rank$ and $select$ are supported in time $O(c)$ \cite{Pat08}.
All these results require the RAM model of computation with $\log n = \Oh{w}$.

\paragraph{Variable-length arrays} An array storing $n$ nonempty strings of 
lengths $l_1,l_2,$ $\ldots,l_n$ can be stored by concatenating the strings and 
adding a bitmap of the same
length of the concatenation, $B = 1\,0^{l_1-1}~1\,0^{l_2-1}\cdots 1\,0^{l_n-1}$.
We can then determine in constant time that
the $i$th string lies between positions $select(B,i)$ and $select(B,i+1)-1$ 
in the concatenated sequence.

\paragraph{Wavelet trees}
A wavelet tree \cite{GGV03} is a binary tree used to represent a sequence
$S[1..n]$, which efficiently supports the queries $access(S,i)$ (the symbol
$S[i]$), $rank_c(S,i)$ (the number of symbols $c$ in $S[1..i]$), and 
$select_c(S,j)$ (the position of the $j$th occurrence of symbol $c$ in $S$).
In this paper we use a wavelet tree variant \cite{BN12} that
uses $n\lg s\,(1+o(1)) + \Oh{s\lg n}$ bits, where the alphabet of $S$ is
$\{1,\ldots,s\}$, and supports the three operations in time 
$\Oh{1+\lg s / \lg w}$.
%Each wavelet tree node corresponds to a subset of the symbols $\{1,\ldots,
%\sigma\}$ of $S$, and implicitly represents the subsequence of $S$ formed by 
%those symbols. The root corresponds to the whole alphabet and implicitly 
%represents the $S$; each leaf represents a singleton $\{c\}$. Each 
%internal node corresponding to subset $X \subseteq \{1,\ldots,\sigma\}$ and 
%representing the subsequence $S_X$ of $S$ has a left child representing a
%subset $X' \subset X$ and a right child representing $X-X'$. The node then 
%stores a bitmap $B_X$ of length $|S_X|$, where $B_X[i]=0$ if the symbol 
%$S_X[i] \in X'$ and so is represented in the left child, otherwise $B_X[i]=1$. 
%With constant-time $rank$ and $select$ operations on the bitmaps (for both $0$s
%and $1$s), operations $access(S,i)$, $rank_c(S,i)$ and $select_c(S,j)$ are 
%solved in time $\Oh{d}$, where $d$ is the depth of the leaf representing 
%$[c,c]$, with $c=S[i]$ in the case of $access$. 
%
%An interesting property of wavelet trees is that, independently of the shape 
%we give them by choosing the subsets corresponding to the nodes, all the 
%involved bitmaps can always be represented in $n\log\sigma\,(1+o(1))$ bits if we
%use a compressed representation (like the one we described) for the bitmaps 
%$B_X$ \cite{GGV03,Nav14}. In addition, we need $\Oh{\sigma\log n}$ bits for 
%the tree pointers.

\subsection{Prefix-free codes}
\label{sec:prefix}

A {\em prefix-free code} (or instantaneous code) is a mapping from a {\em 
source alphabet}, of size $\sigma$, to a sequence of bits, so that each source 
symbol is assigned a {\em codeword} in a way that no codeword is a prefix of 
any other. A sequence of source symbols is then encoded as a sequence of bits 
by replacing each source symbol by its codeword. Compression can be obtained 
by assigning shorter codewords to more frequent symbols \cite[Ch.\ 5]{CT06}.
When the code is prefix-free, we can unambiguously determine each original 
symbol from the concatenated binary sequence, as soon as the last bit of the
symbol's codeword is read. An {\em optimal} prefix-free code minimizes the 
length of the binary sequence and can be obtained with the Huffman algorithm 
\cite{Huf52}.

For constant-time encoding, we can just store a table of $\sigma L$ bits, where 
$L$ is the maximum codeword length, where the codeword of each source symbol is
stored explicitly using standard bit manipulation of computer words
\cite[Sec.\ 3.1]{Nav16}. Since $L=\Oh{w}$, we have to write only $\Oh{1}$
words per symbol. Decoding is a bit less trivial. The classical solution for
decoding a prefix-free code is to store a binary tree $\B$, where each leaf 
corresponds to a source symbol and each root-to-leaf path spells the codeword 
of the leaf, if we write a $0$ whenever we go left and a $1$ whenever we go 
right. Unless the code is obviously suboptimal, every internal node of $\B$ has
two children and thus $\B$ has $\Oh{\sigma}$ nodes. Therefore, it can be 
represented in $\Oh{\sigma\log\sigma}$ bits, which also includes the space to 
store the source symbols assigned to the leaves. By traversing $\B$ from the 
root and following left or right as we read a $0$ or a $1$, respectively, we 
arrive in $\Oh{\ell}$ time at the leaf storing the symbol that is encoded with 
$\ell$ bits in the binary sequence. 

Since $\log\sigma \le L < \sigma$, the above classical solution takes 
$\Oh{\sigma L}$ bits of space. We can reduce the space to 
$\Oh{\sigma\log\sigma}$ bits by deleting the encoding table and adding instead
parent pointers to $\B$, so that from any leaf we can extract the
corresponding codeword in reverse order. Both encoding and decoding take
$\Oh{\ell}$ time in this case.

\begin{figure}[t]
\begin{center}
\includegraphics[width=0.8\textwidth]{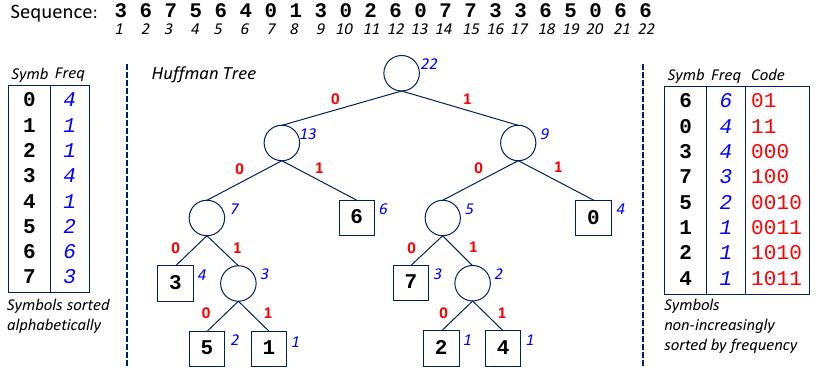}
\end{center}
\vspace*{-5mm}
\caption{An example of Huffman coding. A sequence of symbols on top, the 
symbol frequencies on the left, the Huffman tree $\B$ in the center, and
the corresponding codewords on the right. The blue numbers on the tree nodes
show the total frequencies in the subtrees. The sequence uses $n\lg\sigma = 66$
bits in plain form, but $61$ bits in Huffman-compressed form.}
\label{fig:huffman}
\end{figure}

Figure~\ref{fig:huffman} shows an example of Huffman coding.

\subsection{Canonical prefix-free codes}

By the Kraft Inequality~\cite{Kra49}, we can put any prefix-free code into 
{\em canonical form} \cite{SK64} while maintaining all the codeword lengths.
In the canonical form, the leaves of lower depth are always to the left of 
leaves of higher depth, and leaves of the same depth respect the lexicographic 
order of the source symbols, left to right. 

Canonical codes enable faster encoding and decoding, and/or lower space usage.
Moffat and Turpin \cite{MT97} give practical data structures that can encode
and decode a codeword of $\ell$ bits in time $\Oh{\log\ell}$. Apart from the
$\Oh{\sigma\log\sigma}$ bits they use to store the symbols at the leaves, they 
need $\Oh{L^2}$ bits for encoding and decoding; they do not store the
binary tree $\B$ explicitly. They use the $\Oh{\sigma\log\sigma}$ bits
to map from a symbol $c$ to its left-to-right leaf position $p$ 
and back. Given the increasing positions and codewords of the leftmost leaves
of each length, they find the codeword of a given leaf position $p$ by finding 
the predecessor position $p'$ of $p$, and adding $p-p'$ to the codeword of
$p'$, interpreted as a binary number. For decoding, they extend all those 
first codewords of each length to length $L$, by padding them with $0$s on 
their right. Then, interpreting the first $L$ bits of the encoded stream
as a number $x$, they find the predecessor $x'$ of $x$ among the
padded codewords, corresponding to leaf position $p'$. The leaf position of the
encoded source symbol is then $p'+(x-x')/2^{L-\ell}$, where $\ell$ is the 
depth of the leaf $p$. This is also used to advance by $\ell$ bits in the 
encoded sequence. The time $\Oh{\log\ell}$ is obtained with exponential 
search (binary search would yield $\Oh{\log L}$); the other predecessor time complexities also hold.  

\begin{figure}[t]
\begin{center}
\includegraphics[width=0.8\textwidth]{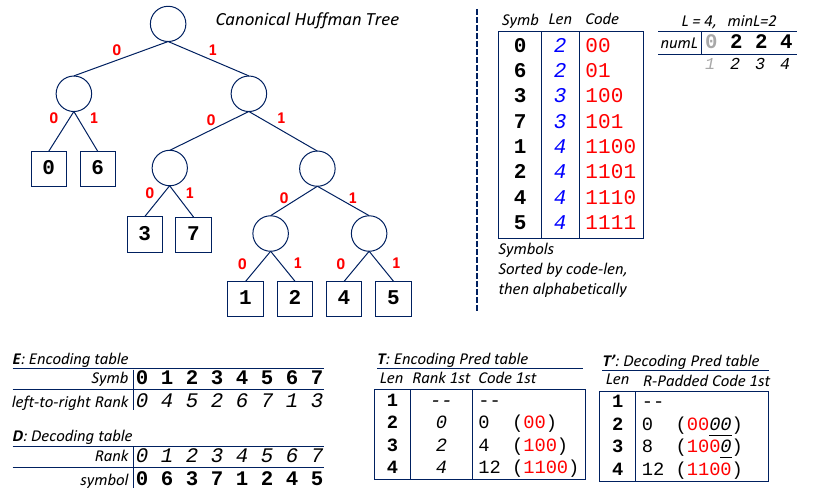}
\end{center}
\vspace*{-5mm}
\caption{The canonical code corresponding to Figure~\ref{fig:huffman}. To
encode a symbol, the table $E$ gives its leaf rank $p$, whose predecessor $p'$
we find in the ranks of table $T$, together with its length $\ell$. We then add
$p-p'$ to the codeword associated with $p'$. To decode $x$, a predecessor 
search for $x$ on the padded codewords of $T'$ finds $x'$. Its associated 
length $\ell$ and leaf position $p'$ are in $T$. We use them 
to obtain the entry in $D$ storing the symbol.}
\label{fig:canonical}
\end{figure}

Figure~\ref{fig:canonical} continues our example with a canonical Huffman code.

Gagie et al.~\cite{GNNO15} improve upon this scheme both in space and time,
by using more sophisticated data structures. They show that, using 
$\Oh{\sigma\log L + L^2}$ bits of space,
constant-time encoding and decoding is possible.

\subsection{Alphabetic codes}
\label{sec:defalpha}

A prefix-free code is {\em alphabetic} if the codewords (regarded as binary 
strings) maintain the lexicographic order of the corresponding source symbols. 
If we build the binary tree $\B$ of such a code, the leaves enumerate the source
symbols in order, left to right. Hu and Tucker \cite{HT71} showed how to build 
an optimal alphabetic code, whose codewords are at most one bit longer than the
optimal prefix-free codes on average \cite{CT06}.

\begin{figure}[t]
\begin{center}
\includegraphics[width=0.8\textwidth]{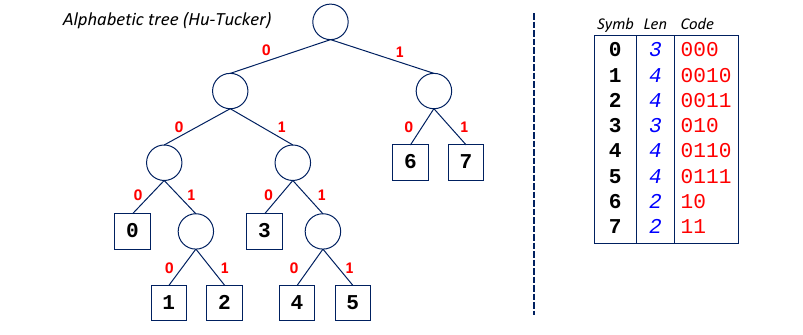}
\end{center}
\vspace*{-5mm}
\caption{An alphabetic code corresponding to the frequencies of 
Figure~\ref{fig:huffman}. The compressed sequence is $62$ bits long.}
\label{fig:alpha}
\end{figure}

Figure~\ref{fig:alpha} gives an alphabetic code tree for our running example.

In an alphabetic code we do not need to map from symbols to leaf positions,
so the sheer topology of $\B$ is sufficient to describe the code.
Such a topology can be described in $\Oh{\sigma}$ bits, in a way that the
tree navigation operations can be simulated in constant time, as well as 
obtaining the left-to-right position of a given leaf and vice versa \cite{MR01}.
With such a representation, we can then simulate the $\Oh{\ell}$ encoding and
decoding algorithms described in Section~\ref{sec:prefix} \cite{GNNO15}.

On the other hand, there is no such a thing like a canonical alphabetic code,
because the leaf left-to-right order cannot be altered. Indeed, no faster
encoding and decoding algorithms exist for alphabetic codes. Our first 
contribution, in Sections~\ref{sec:alphabetic} and \ref{sec:alphabetic2}, 
is a data structure of
$\Oh{\sigma\log L}$ bits that encodes and decodes in time 
$\Oh{\min(\ell,\log L)}$, and even $\Oh{\log\ell}$ if we spend 
$\Oh{2^{L^\epsilon}}$ further bits, for any constant $\epsilon>0$.
While this increases the space compared to the $\Oh{\sigma}$-bit basic 
structure, we show that $o(\sigma)$ bits of space are sufficient to encode
and decode in constant time, if we let the average codeword length increase
by a factor of $1+\Oh{1/\sqrt{\log\sigma}}$ over the optimal.

\subsection{Codes for wavelet matrices}
\label{sec:defwm}

Claude et al.\ \cite{CNO15} showed how to build an optimal prefix-free code 
such that all the codewords of length $\ell$ come before the prefixes of
length $\ell$ of longer codewords in the lexicographic order of the reversed
binary strings.
Specifically, they first build a classical Huffman code and then use the 
Kraft Inequality to build another code with the same codeword lengths and 
with the desired property. They store an $\Oh{\sigma L}$-bit mapping between 
symbols and their codewords, which allows them to encode and decode 
codewords of length $\ell$ in time $\Oh{\ell}$. They use such codes to 
compress wavelet matrices, which are data structures aimed to represent
sequences on large alphabets. Thus, it is worthwhile to devise more space 
economical codeword representations. 

\begin{figure}[t]
\begin{center}
\includegraphics[width=0.8\textwidth]{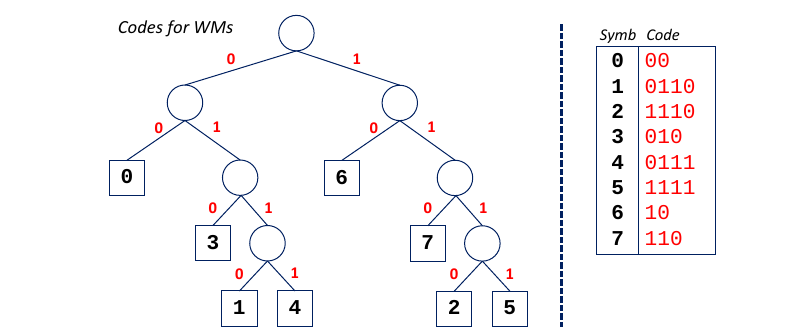}
\end{center}
\vspace*{-5mm}
\caption{A code for wavelet matrices corresponding to the frequencies of 
Figure~\ref{fig:huffman}.}
\label{fig:wmm}
\end{figure}

Figure~\ref{fig:wmm} gives a code tree of this type for our running example.

Our second contribution, in Section~\ref{sec:matrices}, is a representation
for these codes that uses $\Oh{\sigma\log L}$ bits, with the same $\Oh{\ell}$ 
encoding and decoding time. With $\Oh{2^{\epsilon L}}$ further bits, for any 
constant $\epsilon>0$, we achieve constant encoding and decoding time.

\section{Optimal Alphabetic Codes}
\label{sec:alphabetic}

%Katona and Nemetz~\cite{KN76} showed that the longest codeword in a Huffman code for a distribution with minimum positive probability $p_{\min}$ is $\Oh{\log (1 / p_{\min})}$, and Baryan and Kaplan~\cite{BK06} showed this holds for optimal alphabetic codes as well.  Therefore, if we are considering an optimal alphabetic code for a multiset of $n$ elements or a string of length $n$, in the word RAM model we can assume the longest codeword is $\Oh{\log n}$ and therefore fits in a constant number of machine words.

In this section we consider how to efficiently store alphabetic (prefix-free)
codes; recall Section~\ref{sec:defalpha}.
We describe a structure called BSD \cite{GHSV07}, and then how 
we use it to build our fast and compact data structures to store optimal 
alphabetic codes. We finally show how to make it faster using more space.

\subsection{Binary Searchable Dictionaries (BSD)}

Gupta et al.~\cite{GHSV07} describe a structure called {\em BSD}, which encodes
$n$ binary strings of length $L$ using a trie that is analogous to the binary 
tree $\B$ we described above to store the code (except that here all the strings
have the same length $L$). Let us say that the identifier of a string is its 
lexicographic position, that is, the left-to-right position of its leaf in the
trie. Their structure supports extraction of the $i$th string (which is 
equivalent to our encoding), and fast computation of the identifier of a given 
string (which is equivalent to our decoding), both in $\Oh{\log n}$ time.

To achieve this, Gupta et al.\ define a complete binary search tree $T$ on the 
strings with lexicographic order (do not confuse $T$ with the binary trie;
there is one node in $T$ per trie leaf). The complete tree can be stored 
without pointers. Each node $v$ of $T$ represents a string $v.x$, which is not 
explicitly stored. Instead, it stores a suffix $v.t = v.x[l+1..L]$, 
where $l$ is the length of the longest prefix $v.x$ shares with some $u.x$, 
over the ancestors $u$ of $v$ in $T$. For the root $v$ of $T$ it holds that
$v.x = v.t$.

For both operations, we descend in $T$ until reaching the desired node.
We start at the root $v$ of $T$, where we know $v.x$. The invariant is that, as
we descend, we know $v.x$ for the current node $v$ and $u.x$ for all of its 
ancestors $u$ in $T$ (which we have traversed).
Further, we keep track of the most recent ancestors $u.l$ and $u.r$ from where 
our path went to the left and to the right, respectively, and therefore it 
holds that $u=u_l$ if $v.t[1]=0$ and $u=u_r$ if $v.t[1]=1$ \cite{GHSV07}.
Whenever we choose the child $v'$ of $v$ to follow, we compute
$v'.x$ by composing $v'.x = u.x[1..L-|v'.t|] \cdot v'.t$, which restores 
the invariant. The procedure ends after $\Oh{\log n}$ constant-time steps, and
we can do the concatenation that computes $v'.x$ in constant time in the RAM 
model.

To extract the $i$th string, we navigate from the root towards the $i$th node
of $T$. Because $T$ is a complete binary search tree, we know algebraically 
whether the $i$-th node is $v$, or it is to the left or to the right of $v$. 
If it is $v$, we already know $v.x$, as explained, and we are done. Otherwise, 
we choose the proper child $v'$ of $v$ and continue the search.
Finding $i$ from its string $x$ is analogous, except that we compare $x$ with
$v.x$ numerically (in constant time in the RAM model) to determine whether we
have found $v$ or we must go left or right. Because $T$ is complete, we know
algebraically the identifier $v.i$ of each node $v$ without need of storing it.

Gupta et al.~\cite{GHSV07} show that, surprisingly, the sum of the lengths of 
all the strings $v.t$ is bounded by the number of edges in the trie.
Our data structure for optimal alphabetic codes builds on this BSD data 
structure.

\subsection{Our data structure}

Given an optimal alphabetic code over a source alphabet of size $\sigma$ with 
maximum codeword length $L$, we store the lengths of the $\sigma$ codewords 
using $\sigma\lceil \log L\rceil$ bits, and then pad the codewords on the right
with 0s
up to length $L$.  We divide the lexicographically sorted padded codewords into
blocks of size $L$ (the last block may be smaller). We collect the first padded
codeword of every block in a predecessor data structure, and store all the 
(non-padded) codewords of each block in a BSD data structure, one per block.

The predecessor data structure then stores $\lceil \sigma/L \rceil$ numbers
in a universe of size $2^L$. As seen in Section~\ref{sec:basics}, the structure
uses $\Oh{(\sigma/L)\log(2^L)} = \Oh{\sigma}$ bits and
answers predecessor queries in time $\Oh{\log\log(2^L)} = \Oh{\log L}$.

Each BSD structure, on the other hand, stores (at most) $L$ strings $v.t$.
Unlike the original BSD structure, our codewords are of varying length
(those lengths were stored separately, as indicated). This does not invalidate
the argument that the sum of the strings $v.t$ adds up to the number of edges in
the trie of the $L$ codewords: what Gupta et al.~\cite[Lem.~3]{GHSV07} show
is that each edge of the trie is mentioned in only one string $v.t$, with no
reference to the code lengths. 
We vary its encoding, though: We store all the strings $v.t$ of the BSD, 
in the same order of the nodes of $T$, concatenated in a variable-length array 
as described in Section~\ref{sec:basics}. With constant-time $select$ we find 
where is $v.t$ in the concatenation, and with another $\Oh{1}$ time we extract
it in the RAM model. 

Considering the extra space needed to find in constant time where is $v.t$, we 
spend $\Oh{1}$ bits per trie edge. Since the trie stores up to $L$ consecutive
leaves of the whole binary tree $\B$ (and internal nodes of $\B$ have two
children because the alphabetic code is optimal), it follows that the trie
has $\Oh{L}$ nodes: There are $\Oh{L}$ trie nodes with two children because
there are $L$ leaves in the trie, and the trie nodes with one child are those
leading to the leftmost and rightmost trie leaves. Since the leaves are of 
depth $L$, there are $\Oh{L}$ of those trie nodes too.
Therefore, we use $\Oh{L}$ bits per BSD structure,
adding up to $\Oh{\sigma}$ bits overall. 

The total space is then dominated by the $\sigma\log L + \Oh{\sigma}$ bits 
spent in
storing the lengths of the codewords. On top of that, the predecessor data 
structure uses $\Oh{\sigma}$ bits and the BSD structures use other $\Oh{\sigma}$
bits.

To encode symbol $i$, we go to the $\lceil i/L\rceil$th BSD structure and
find the $i'$th string inside it, with $i'=i-(\lceil i/L\rceil-1)\cdot i$.
The algorithm is identical to that for BSD, except that each $v.x$ has 
variable length; recall that we have those lengths $|v.x|$ stored explicitly. 
We thus update $v'.x = u.x[1..|v'.x|-|v'.t|] \cdot v'.t$ when moving to node
$v'$.

To decode, we store in a number $x$ the first $L$ bits of the stream, find its 
predecessor in our structure, and decode $x$ in the corresponding BSD structure.
The only difference is that, when we compare $x$ with $v.x$, their lengths
differ (because we do not know the length $\ell$ of the codeword we seek, which
prefixes $x$). Since the code is prefix-free, it follows that the codeword we
look for is $v.x$ if $v.x = x[1..|v.x|]$, otherwise we go left or right
according to which is smaller between those $|v.x|$-bit numbers. When we find
the proper node $v$, the source symbol is the position $i$ of $v$ (which we
compute algebraically, as explained) and the length of the codeword is 
$\ell = |v.x|$.

In both cases, the time is $\Oh{\log L}$ to find the proper node in the BSD
plus, in the case of decoding, $\Oh{\log L}$ time for the predecessor search.
As before, we can also encode and decode a codeword of length $\ell$ in time 
$\Oh{\ell}$ using the basic $\Oh{\sigma}$-bit representation. We can even choose
the smallest by attempting the encoding/decoding up to $\log L$ steps, and then
switch to the $\Oh{\log L}$-time procedure if we have not yet finished.

\begin{theorem}
\label{thm:optimal}
Given a probability distribution over an alphabet of $\sigma$ symbols, we 
can build an optimal alphabetic prefix-free code and store it in 
$\sigma \log L + \Oh{\sigma}$ bits, where $L$ is the maximum codeword length, 
such that we can encode and decode any codeword of length $\ell$ in 
$\Oh{\min (\ell, \log L)}$ time.
The result assumes a $w$-bit RAM computation model with $L=\Oh{w}$.
\end{theorem}

\begin{figure}[t]
\begin{center}
\includegraphics[width=\textwidth]{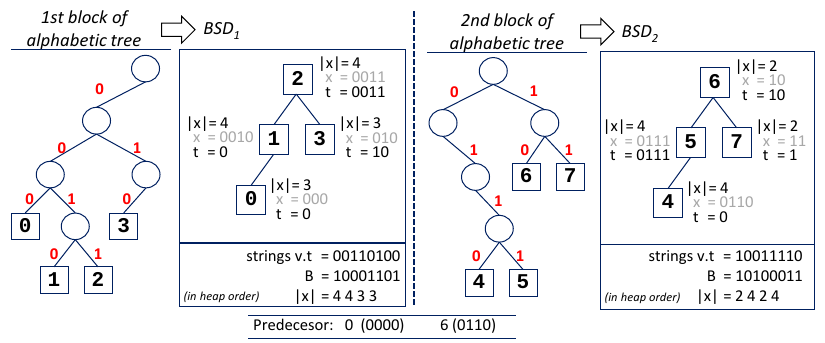}
\end{center}
\vspace*{-5mm}
\caption{Our representation of the code for wavelet matrices of 
Figure~\ref{fig:wmm}. For each BSD structure we only store the
concatenated strings $v.t$, their bitmap $B$, and the code lengths $|x|$.
The first codes of each BSD structure are stored in the predecessor structure
on the bottom, padded to $L=4$ bits.}
\label{fig:bsd}
\end{figure}

Figure~\ref{fig:bsd} shows our structure for the codewords tree of
Figure~\ref{fig:wmm}. Note that, for each BSD structure, the length of the
concatenated strings $v.t$ equals the number of edges in the corresponding
piece of the codewords tree. For example, to encode the symbol {\bf 3}, we
must encode the 4th symbol of $BSD_1$. We start at the root $u$ (corresponding
to symbol {\bf 2}), with $u.x=u.t=0011$. We know algebraically that the root
corresponds to the 3rd symbol, so we go right to $v$, the node representing
the symbol {\bf 3}. Since $v.t[1]=1$, $v.t$ is encoded with respect to the
nearest ancestor where we went right, that is, from the root $u$. We have
$|v.x|=3$ stored explicitly, so we build $v.x = u.x[1..|v.x|-|v.t|] \cdot v.t
= 0\cdot 10$. Since we know algebraically that we arrived at the 4th symbol, 
we are done: the codeword for {\bf 3} is $010$. Let us now decode $0110=6$.
The predecessor search tells it appears in $BSD_2$. We start at the root $u$
(which encodes {\bf 6}). Since its extended codeword, $u.x = 10\cdot 00$, is
larger than $0110$, we go left to the node $v$ that represents {\bf 5}. Since
$v.t[1]=0$, $v.t$ is represented with respect to the last ancestor where we
went left, that is, $u$. So we compose $v.x = u.x[1..|v.x|-|v.t|] \cdot v.t =
\cdot 0111$. Now, since $v.x = 0111$ is larger than our codeword $0110$, we 
again
go left to the node $v'$ that represents {\bf 4}. Since $v'.t[1]=0$, $v'.t$
is also represented with respect to the last node where we went left, that is,
$v.x$. So we compose $v'.x = v.x[1..|v'.x|-|v'.t|] \cdot v'.t = 011 \cdot 0$.
We have found the code sought, $0110$, and we algebraically know that the node
corresponds to the source symbol {\bf 4}.

\subsection{Faster operations}

In order to reduce the time $\Oh{\min(\ell,\log L)}$ to $\Oh{\log\ell}$, we 
manage to encode and decode in constant time the codewords of length up to
$L' = L^{\epsilon/2}$, for some constant $\epsilon>0$. For the longer codewords,
since $L' < \ell \le L$, it holds that $\log\ell = \Theta(\log L)$, and thus
we already process them in time $\Oh{\log\ell}$.

For encoding, we store a bitmap $B[1..\sigma]$, so that $B[i]=1$ iff the
length of the codeword of the $i$th source symbol is at most $L'$. We also 
store a table $S[1..2^{L'}]$ so that, if $B[i]=1$, then $S[rank(B,i)]$ stores 
the codeword of the $i$th source symbol (only $2^{L'}$ source symbols can have
codewords of length up to $L'$). To encode $i$, we check
$B[i]$. If $B[i]=1$, then we output the codeword $S[rank(B,i)]$ in constant
time; otherwise we encode $i$ as in Theorem~\ref{thm:optimal}

For decoding, we build a table $A[0..2^{L'}-1]$ where, for any 
$0 \le j < 2^{L'}$, if the binary representation of $j$ is prefixed by the 
codeword of the $i$th codeword, which is of length $\ell \le L'$, then
$S[j] = (i,\ell)$. Instead, if no codeword prefixes $j$, then $S[j]=\perp$.
We then read the next $L$ bits of the stream and extract the first $L'$ of those
$L$ bits in a number $j$. If $S[j]=(i,\ell)$, then we have decoded the symbol
$i$ in constant time and advance in the stream by $\ell$ bits. Otherwise, we
proceed with the $L$ bits we have read as in Theorem~\ref{thm:optimal}.

The encoding and decoding time is then always bounded by $\Oh{\log\ell}$, as 
explained. The space for $B$, $S$, and $A$ is 
$\Oh{\sigma + 2^{L'}(L'+\log\sigma)} \subseteq
\Oh{\sigma + 2^{L^\epsilon}}$ bits, because $L'+\log\sigma = \Oh{L}$ and
$\Oh{L2^{L^{\epsilon/2}}} \subseteq 2^{L^\epsilon}$.

\begin{corollary}
\label{cor:optimal}
Given a probability distribution over an alphabet of $\sigma$ symbols, we can 
build an optimal alphabetic prefix-free code and store it in $\Oh{\sigma \lg L+
2^{L^\epsilon}}$ bits, where $L$ is the maximum codeword length and $\epsilon$ 
is any positive constant, such that we can encode and decode any codeword of 
length $\ell$ in $\Oh{\log \ell}$ time.
The result assumes a $w$-bit RAM computation model with $L=\Oh{w}$.
\end{corollary}

\section{Near-Optimal Alphabetic Codes}
\label{sec:alphabetic2}

Our approach to storing a nearly optimal alphabetic code compactly has two parts: first, we show that we can build such a code so that the expected codeword length is \(\left(1 + \Oh{1 / \sqrt{\log \sigma}}\right)^2 = 1 + \Oh{1 / \sqrt{\log \sigma}}\) times the optimal, the codewords tree $\B$ has height at most \(\lg \sigma + \sqrt{\lg \sigma} + 3\), and each subtree rooted at depth \(\lceil \lg \sigma  - \sqrt{\lg \sigma} \rceil\) is completely balanced. Then, we manage to store such a tree in \(o (\sigma)\) bits so that encoding and decoding take $\Oh{1}$ time.

\subsection{Balancing the codewords tree}

Evans and Kirkpatrick~\cite{EK04} showed how, given a binary tree on $\sigma$ leaves, we can build a new binary tree of height at most \(\lceil \lg \sigma \rceil + 1\) on the same leaves in the same left-to-right order, such that the depth of each leaf in the new tree is at most 1 greater than its depth in the original tree. We can use their result to restrict the maximum codeword length of an optimal alphabetic code, for an alphabet of $\sigma$ symbols, to be at most \(\lg \sigma + \sqrt{\lg \sigma} + 3\), while forcing its expected codeword length to increase by at most a factor of \(1 + \Oh{1 / \sqrt{\log \sigma}}\). To do so, we build the tree $\B$ for an optimal alphabetic code and then rebuild, according to Evans and Kirkpatrick's construction, each subtree rooted at depth \(\lceil \sqrt{\lg \sigma} \rceil\).  The resulting tree, $\B_{lim}$, has height at most \(\lceil \sqrt{\lg \sigma} \rceil + \lceil \lg \sigma \rceil + 1\) and any leaf whose depth increases was already at depth at least \(\lceil \sqrt{\lg \sigma} \rceil\).
Although there are better ways to build a tree $\B_{lim}$ with such a height 
limit \cite{Wes76,Ita76}, our construction is sufficient to obtain an 
%There are better ways to build a tree $\B_{lim}$ with such a height limit.
%Itai~\cite{Ita76} and Wessner~\cite{Wes76} independently showed how, given a
%probability distribution over an alphabet of $\sigma$ characters, we can
%build an alphabetic code $T_{lim}$ that has maximum codeword length at
%most \(\lg \sigma + \sqrt{\lg \sigma} + 3\) and is optimal among all such
%codes. Our construction in the previous paragraph, even if not optimal, shows
%that the expected codeword length of $T_{lim}$ is at most \(1 + \Oh{1 /
expected codeword length for $\B_{lim}$ that is \(1 + \Oh{1 /
\sqrt{\log \sigma}}\) times the optimal.

Further, let us take $\B_{lim}$ and completely balance each subtree rooted at depth \(\lceil \lg \sigma  - \sqrt{\lg \sigma} \rceil\).
The height does not increase and any leaf whose depth increases was already at depth at least \(\lceil \lg \sigma - \sqrt{\lg \sigma} \rceil\), so the expected codeword length increases by at most a factor of
\[\frac{\lceil\sqrt{\lg \sigma}\rceil + \lceil\lg\sigma\rceil + 1}{\lceil \lg \sigma - \sqrt{\lg \sigma} \rceil}
= 1 + \Oh{1 / \sqrt{\log \sigma}}\,.\]
Let $\B_{bal}$ be the resulting tree.  Since the expected codeword length of $\B_{lim}$ is in turn a factor of \(1 + \Oh{1 / \sqrt{\log \sigma}}\) larger than that of $\B$, the expected codeword length of $\B_{bal}$ is also a factor of \(\left(1 + \Oh{1 / \sqrt{\log \sigma}}\right)^2 = 1 + \Oh{1 / \sqrt{\log \sigma}}\) larger than
the optimal. The tree $\B_{bal}$ then describes our suboptimal code.

\subsection{Representing the balanced tree}

To represent $\B_{bal}$, we store a bitmap \(B [1..\sigma]\) in which \(B [i] = 1\) if and only if the $i$th left-to-right leaf is:
\begin{itemize}
\item of depth less than \(\lceil \lg \sigma  - \sqrt{\lg \sigma} \rceil\), or 
\item the leftmost leaf in a subtree rooted at depth \(\lceil \lg \sigma  - \sqrt{\lg \sigma} \rceil\).  
\end{itemize}
Note that each $1$ of $B$ corresponds to a node of $\B_{bal}$ with depth at most
$\lceil \lg \sigma  - \sqrt{\lg \sigma} \rceil$. Since there are
$m=\Oh{2^{\lg \sigma  - \sqrt{\lg \sigma}}}$ such nodes, $B$ can be 
represented in compressed form as described in Section~\ref{sec:basics}, using
$m\log(\sigma/m)+\Oh{m+\sigma/\log^c \sigma} = 
\Oh{2^{\lg \sigma-\sqrt{\lg \sigma}} \lg\sigma + \sigma/\log^c \sigma}$ bits,
supporting $rank$ and $select$ in time $\Oh{c}$. For any constant $c$, the
term $\Oh{2^{\lg \sigma-\sqrt{\lg \sigma}} \lg\sigma} = 
\Oh{\sigma / 2^{\sqrt{\lg \sigma}-\lg\lg\sigma}}$ is dominated by the second
component, $\Oh{\sigma/\log^c \sigma}$.

%Let us for simplicity assume that the alphabet is $[1..\sigma]$.
For encoding in constant time we store an array \(S[1..2^{\lceil \lg \sigma  - \sqrt{\lg \sigma} \rceil}]\), which explicitly stores the codewords assigned to 
the leaves of $\B_{bal}$ where $B[i]=1$, in the same order of $B$. That is, if
$B[i]=1$, then the code assigned to the symbol $i$ is stored at $S[rank(B,i)]$. Since the codewords are of length at most $\lceil\sqrt{\lg\sigma}\rceil+
\lceil\lg\sigma\rceil+1=\Oh{\log\sigma}$, $S$ requires 
\(\Oh{2^{\lg \sigma - \sqrt{\lg \sigma}} \log \sigma} = o(\sigma / \log^c \sigma)\) bits of space, for any constant $c$. We can
also store the length of the code within the same asymptotic space.

To encode the symbol $i$, we check whether \(B [i] = 1\) and, if so, we simply look up the codeword in $S$ as explained.  If \(B [i] = 0\), we find the preceding 1 at $i'=select(B,k)$ with $k=rank(B,i)$, which marks the leftmost leaf in the subtree rooted at depth \(\lceil \lg \sigma  - \sqrt{\lg \sigma} \rceil\) that contains the $i$th leaf in $\B$. Since the subtree is completely balanced, we can compute the code for the symbol $i$ in constant time from that of the symbol $i'$: The balanced subtree has $r=i''-i'$ leaves, where
$i'' = select(B,k+1)$, and its height is $h=\lceil \lg r\rceil$. Then
the first $2r-2^h$ codewords are of the same length of the codeword for $i'$,
and the last $2^h-r$ have one bit less. Thus, if $i-i' < 2r-2^h$, the codeword
for $i'$ is $S[k]+i-i'$, of the same length of that of $i$; otherwise
it is one bit shorter, $(S[k]+2r-2^h)/2+i-i'-(2r-2^h) =
S[k]/2+i-i'-(r-2^{h-1})$.

To be able to decode quickly, we store an array \(A [0..2^{\lceil \lg \sigma  - \sqrt{\lg \sigma} \rceil}-1]\) such that, if the \(\lceil \lg \sigma  - \sqrt{\lg \sigma} \rceil\)-bit binary representation of \(j\) is prefixed by the $i$th codeword, then \(A [j]\) stores $i$ and the length of that codeword. If, instead, the \(\lceil \lg \sigma  - \sqrt{\lg \sigma} \rceil\)-bit binary representation of $j$ is the path label to the root of a subtree of $\B_{bal}$ with size more than 1, then \(A [j]\) stores the position $i'$ in $B$ of the leftmost leaf in that subtree (thus $B[i']=1$).  Again, $A$ takes \(\Oh{2^{\log \sigma - \sqrt{\log \sigma}} \log \sigma} = o(\sigma / \log^c \sigma)\) bits for any constant $c$.

Given a string prefixed by the $i$th codeword, we take the prefix of length \(\lceil \lg \sigma  - \sqrt{\lg \sigma} \rceil\) of that string (padding with 0s on the right if necessary), view it as the binary representation of a number $j$, and check \(A [j]\).  This either tells us immediately $i$ and the length of the $i$th codeword, or tells us the position $i'$ in $B$ of the leftmost leaf in the subtree containing the desired leaf. In the latter case, since the subtree is completely balanced, we can compute $i$ in constant time: We find $i''$, $r$,
and $h$ as done for encoding. We then take the first $\lceil \lg\sigma - 
\sqrt{\lg\sigma}\rceil+h$ bits of the string
(including the prefix we had already read, and padding with a 0 if necessary), and interpret it as the number $j'$. Then, if
$d = j'-S[rank(B,i')] < 2r-2^h$, it holds $i = i'+d$. Otherwise, the code is 
one bit shorter and the decoded symbol is $i= i'+2r-2^h + \lfloor (d-(2r-2^h))/2
\rfloor = i'+r-2^{h-1}+\lfloor d/2 \rfloor$.

\begin{figure}[t]
\begin{center}
\includegraphics[width=0.8\textwidth]{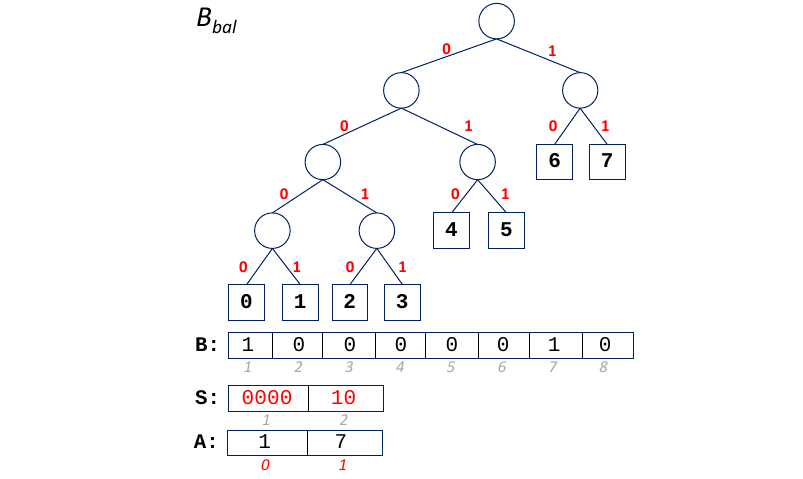}
\end{center}
\vspace*{-5mm}
\caption{The alphabetic tree of Figure~\ref{fig:alpha} balanced from
level $1$. The resulting compressed sequence length is now $67$ bits (larger
than a plain code, in this toy example).}
\label{fig:alphabal}
\end{figure}

Figure~\ref{fig:alphabal} shows an example, where we have balanced from level
$1$ instead of level $2$ (which is what the formulas indicate) so that the
tree of Figure~\ref{fig:alpha} undergoes some change. The subtrees starting
at the two children of the root are then balanced and made complete. The array
$S$ gives the codeword of the first leaves of both subtrees and $A$ gives the
position in bitmap $B$ of the codewords of the nodes rooting the balanced
subtrees. To encode {\bf 2}, since it is the 3rd symbol ($i=3$), we compute
$k=rank(B,3)=1$, $i'=select(B,1)=1$, $i'' = select(B,1+1)=7$, and 
$S[1] = 0000$. The complete subtree then has $r=i''-i' = 6$ leaves and its 
height is $r=\lceil \lg 6 \rceil = 3$. The first $2r-2^h = 4$ leaves are of
depth $4$ like $S[1]$, and the other $2^h-r=2$ are of depth $3$. Since
$i-i' = 2 < 4$, our codeword is of length $4$ and is computed as
$S[1]+i-i' = 0010$. Instead, to decode $010$, we truncate it to length $1$,
obtaining $j=0$. Since $A[0]=1$, the code is in the subtree that starts at
$i'=1$ in $B$. We compute $i''=7$, $r=6$, and $h=3$ as before. The first $1+h=4$
bits of our code is $j'=0100$, which we had to pad with a $0$. Since 
$d=j'-S[rank(B,1)]=0100-0000 = 4 \ge 2r-2^h$, the code is of length $3$ and
the source symbol is $i=1+6-2^2+2=5$, that is, {\bf 4}.

\begin{theorem}
\label{thm:nearly_optimal}
Given a probability distribution over an alphabet of $\sigma$ symbols, we can build an alphabetic prefix-free code whose expected codeword length is at most a factor of \(1 + \Oh{1 / \sqrt{\log \sigma}}\) more than optimal and store it in $\Oh{\sigma / \log^c \sigma}$ bits, for any constant $c$, such that we can encode and decode any symbol in constant time $\Oh{c}$.
\end{theorem}

\section{Efficient Codes for Wavelet Matrices}
\label{sec:matrices}

We now show how to efficiently represent the prefix-free codes for wavelet 
matrices; recall Section~\ref{sec:defwm}. We first describe a representation
based on the wavelet trees of Section~\ref{sec:basics}. This is then used to
design a space-efficient version that encodes and decodes codewords of length 
$\ell$ in time $\Oh{\ell}$, and then a larger one that encodes and decodes
in constant time.

\subsection{Using wavelet trees}

Given a code for wavelet matrices, we reassign the codewords of the same length such that the lexicographic order of the reversed codewords of that length is the same as that of their symbols. This preserves the property that the codewords
of some length are numerically smaller than the corresponding prefixes of 
longer codewords in the lexicographic order of their reverses. The positive
aspect of this reassignment is that all the information on the code can be 
represented in $\sigma\lg L$ bits as a sequence $D = d_1, \ldots, d_\sigma$, 
where $d_i$ is the depth of the leaf encoding symbol $i$ in the codewords tree 
$\B$. We can represent $D$ with a wavelet tree using $\sigma\lg L\,(1+o(1)) +
\Oh{L\lg\sigma} \subseteq \Oh{\sigma\lg L}$ bits%
\footnote{Since $L \le \sigma$, $L/\log L \le \sigma/\log\sigma$ because
$x/\lg x$ is increasing for $x \ge 3$, thus $L\lg \sigma \le \sigma\lg L$
for all $3 \le L \le \sigma$ and $\Oh{L\lg\sigma} \subseteq\Oh{\sigma\lg L}$.}
(Section~\ref{sec:basics}), and then:
\begin{itemize}
\item $access(D,i)$ is the length $\ell$ of the codeword of symbol $i$; 
\item $rank_\ell(D,i)$ is the position (in reverse lexicographic order) of the 
leaf representing symbol $i$ among those of codeword length $\ell$; and 
\item $select_\ell(D,r)$ is the symbol corresponding to the $r$th codeword 
of length $\ell$ (in reverse lexicographic order).
\end{itemize}

Those operations take time $\Oh{1+\lg L / \lg w}$, because the 
alphabet of $D$ is $\{1,\ldots,L\}$. Since we assume $L=\Oh{w}$ 
(Section~\ref{sec:assumptions}), this time is $\Oh{1}$.

%In order to perform those operations in time $\Oh{\ell}$, where $\ell$ is the 
%length of the codeword involved in the operation, we can give the wavelet tree 
%of $D$ the same shape of the tree $\B$. That is, the set $X$ corresponding to
%a wavelet tree node is formed by the symbols associated with the leaves 
%descending from the same node in $\B$.
%We can even perform the operations in time $\Oh{\log \ell}$ by using
%a wavelet tree shaped like the trie for the first $\sigma$ codewords represented
%with Elias $\gamma$- or $\delta$-codes~\cite[Observation 1]{GHMN11}. The size stays
%$\Oh{\sigma\log L}$ if we use compressed bitmaps at the nodes
%\cite{GGV03,Nav14}.

We are left with two subproblems. For decoding the first symbol encoded in a binary string, we need to find the length $\ell$ of its codeword and the lexicographic rank $r$ of its reverse among the reversed codewords of that length. With that information we have that the source symbol is $select_\ell(D,r)$. For encoding a symbol $i$, instead, we find the length $\ell=D[i]$ of its codeword and the lexicographic rank $r=rank_\ell(D,i)$ of its reverse among the reversed codewords of length $\ell$. Then we must find the codeword given $\ell$ and $r$. 

We first present a solution that takes $\Oh{L \log \sigma} \subseteq \Oh{\sigma\log L}$ further bits and works in $\Oh{\ell}$ time. We then present a solution that takes $\Oh{2^{\epsilon L}}$ further bits, for any constant $\epsilon>0$,
 and works in less time.

\subsection{A space-efficient representation}

For each depth $d$ between 0 and $L$, let \(\nodes (d)\) be the total number of nodes at depth $d$ in $\B$ and let \(\leaves (d)\) be the number of leaves at depth $d$.  Let $v$ be a node other than the root, let $u$ be $v$'s parent, let $r_v$ be the lexicographic rank (counting from 1) of $v$'s reversed path label among all the reversed path labels of nodes at $v$'s depth, and let $r_u$ be defined analogously for $u$. Then note the following facts:
\begin{enumerate}
\item Because $\B$ is optimal, every internal node has two children, so half the non-root nodes are left children and half are right children.  
\item Because the reversed path labels of the left children at any depth 
start with a $0$, they are all lexicographically less than the reversed path 
labels of all the right children at the same depth, %(or, indeed, at any depth),
which start with a $1$.  
\item Because of the ordering properties of these codes, the reversed path labels of all the leaves at any depth are lexicographically less than the reversed path labels of all the internal nodes at that depth.
\end{enumerate}

It then follows that:
\begin{itemize}
\item $v$ is a leaf if and only if $r_v \le \leaves(\depth(v))$;
\item $v$ is $u$'s left child if and only if \(r_v \leq \nodes (\depth (v)) / 2\);
\item if $v$ is $u$'s left child then \(r_v = r_u - \leaves (\depth (u))\); and
\item if $v$ is $u$'s right child then \(r_v = r_u - \leaves (\depth (u)) + \nodes (\depth (v)) / 2\).
\end{itemize}
Of course, by rearranging terms we can also compute $r_u$ in terms of $r_v$.

We store \(\nodes (d)\) and \(\leaves (d)\) for $d$ between 0 and $L$, which 
requires $\Oh{L\log\sigma}$ bits. With the formulas above, we can decode the
first codeword, of length $\ell$, from a binary string as follows:
We start at the root $u$, $r_u=1$, and descend in $\B$ until we reach the leaf 
$v$ whose path
label is that codeword, and return its depth $\ell$ and the lexicographic rank 
$r=r_v$ of its reverse path label among all the reversed path labels of nodes 
at that depth. We then compute $i$ from $\ell$ and $r$ as described with the
wavelet tree. %, in $\Oh{\ell}$ additional time. 
Note that these nodes $v$ are
conceptual: we do not represent the nodes explicitly, but we still can compute
$r_v$ as we descend left or right; we also know when we have reached a 
conceptual leaf. 

For encoding $i$, we obtain as explained, with the wavelet tree, its length 
$\ell$ and the rank $r=r_v$ of its reversed codeword among the reversed
codewords of that length. Then we use the formulas to walk up towards the
root, finding in each step the rank $r_u$ of the parent $u$ of $v$, and
determining if $v$ is a left or right child of $u$. This yields the $\ell$
bits of the codeword of $i$ in reverse order (0 when $v$ is a left child of
$u$ and 1 otherwise), in overall time $\Oh{\ell}$. This completes our first
solution, which we evaluate experimentally in Section~\ref{sec:exp}.

\begin{theorem}
\label{thm:matrices0}
Consider an optimal prefix-free code in which all the codewords of length 
$\ell$ come before the prefixes of length $\ell$ of longer codewords in the 
lexicographic order of the reversed binary strings.
We can store such a code in $\sigma\lg L\,(1+o(1)) + \Oh{L\lg\sigma} \subseteq \Oh{\sigma \log L}$ bits --- possibly after swapping symbols' codewords of the same length --- where $\sigma$ is the alphabet size and $L$ is the maximum codeword length, so that we can encode and decode any codeword of length $\ell$ in 
$\Oh{\ell}$ time.
The result assumes a $w$-bit RAM computation model with $L=\Oh{w}$.
\end{theorem}

\begin{figure}[t]
\begin{center}
\includegraphics[width=0.8\textwidth]{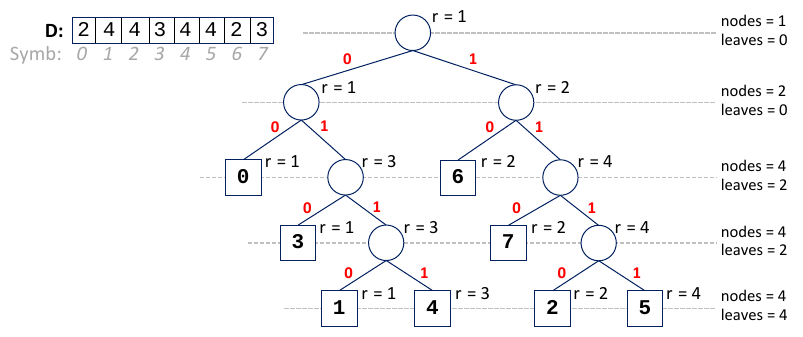}
\end{center}
\vspace*{-5mm}
\caption{Our representation for the tree of Figure~\ref{fig:wmm}. We only
store the sequence $D$ and the values $\nodes$ and $\leaves$ at each level.
For each node $v$ we show its $r_v$ value.}
\label{fig:wmmeffic}
\end{figure}

Figure~\ref{fig:wmmeffic} shows our representation for the codewords tree of
Figure~\ref{fig:wmm}. To decode $110...$, we start at the root with $r_0=1$.
The next bit to decode is a $1$, so we must go right: the node of depth $1$ is
then $r_1 = r_0-\leaves(0)+\nodes(1)/2 = 2$. The next bit to decode is again
a $1$, so we go right again: the node of depth $2$ is $r_2 = r_1 - \leaves(1)+
\nodes(2)/2 = 4$. The last bit to decode is a $0$, so we go left: the node of
depth $3$ is $r_3 = r_2 - \leaves(2) = 2$. Now we are at a leaf (because
$r_3 \le \leaves(3)=2$) whose depth is $\ell=3$ and its rank is $r=r_3=2$. The 
corresponding symbol is then $select_3(D,2)=8$, that is, symbol {\bf 7}. 
Instead, to encode {\bf 3}, the symbol number $i=4$, we compute its codeword 
length $\ell=D[4]=3$ and its rank $r=rank_3(D,4)=1$. Our leaf then corresponds 
to $r_3=1$, and we discover the code in reverse order by waking upwards to the 
root. Since $r_3 \le \nodes(3)/2=2$, we are a left child (so the codeword ends
with a $0$) and our parent has $r_2 = r_3 + \leaves(2) = 3$. Since $r_2 >
\nodes(2)/2 = 2$, this node is a right child (so the codeword ends with $10$)
and its parent has $r_1 = r_2+\leaves(1)-\nodes(2)/2 = 1$. Finally, the new
node is a left child because $r_1 \le \nodes(1)/2=1$, and therefore the
codeword is $010$.

\begin{figure}[t]
\begin{center}
\includegraphics[width=0.8\textwidth]{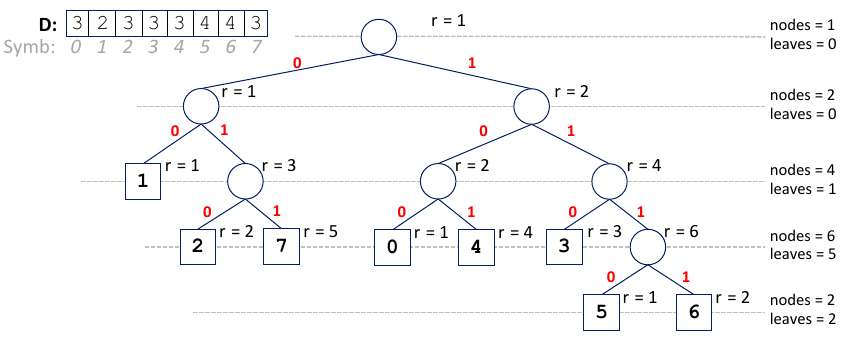}
\end{center}
\vspace*{-5mm}
\caption{The representation of a less regular code, with the same notation
of Figure~\ref{fig:wmmeffic}, produced for the sequence ``$14765232100214171$''.}
\label{fig:wmmeffic2}
\end{figure}

Figure~\ref{fig:wmmeffic2} shows another example with a sequence producing
a less regular tree.
Consider decoding $1110...$. We start at the root with $r_0=1$.
The first bit to decode is a $1$, so we go right and obtain 
$r_1 = r_0-\leaves(0)+\nodes(1)/2 = 2$. The next bit is also
a $1$, so we go right again and get $r_2 = r_1 - \leaves(1)+
\nodes(2)/2 = 4$. The third bit to decode is also a $1$, so we 
go right again to get $r_3 = r_2 - \leaves(2)+ \nodes(3)/2 = 6$ 
(that is, the $4$th node of level $2$, minus the leaf with code $00$,
shifted by all the $6/3$ nodes of level $3$ that descend by a $0$ 
and thus precede our node). Finally, the next bit is a $0$, so we
go left, to node $r_4 = r_3 - \leaves(3) = 1$ (that is, the $6$th 
node of level $3$ minus the $5$ leaves of that level). Now we are 
at a leaf because $r_4 \le \leaves(4)=2$. We leave to the reader 
finding the corresponding symbol {\bf 5} in $D$, as done for the
previous example, as well as working out the decoding of the same 
symbol.

\subsection{Faster and larger}

We now show how to speed up the preceding procedure so that we 
can perform $t$ steps on the tree in constant time, for some given $t$.
From the formulas that relate $r_u$ and $r_v$ it is apparent that, given a node
$u$ and the following $t$ bits to decode, the node $x$ we will arrive at 
depends only on the $\nodes$ and $\leaves$ values at the depths
$\depth(u),\ldots,\depth(u)+t$. More precisely, the value $r_x$ 
is $r_u$ plus a number that depends only on the involved 
depths and the $t$ bits of the codeword to decode. Similarly, given $r_x$,
the last $t$ bits leading to it, and the rank $r_u$ of the ancestor $u$ of
$x$ at distance $t$, depend on the same values of $\nodes$ and $\leaves$.

Let us first consider encoding a source symbol. We obtain its codeword length
$\ell$ and rank $r$ from the wavelet tree, and then extract the codeword.
Consider all the path labels of a particular length that end with a particular 
suffix of length $t$: the lexicographic ranks of their reverses are consecutive,
forming an interval. We can then partition the nodes at any depth $d$ by 
those intervals of rank values. 

Let $x$ be a node at depth $d$, $u$ be its ancestor at distance $t$, and $r_x$ 
and $r_u$ be the rank values of $x$ and $u$, respectively. As per the previous
paragraph, the partition interval where $r_x$ lies determines the last $t$ bits
of $x$'s path label, and it also determines the difference between $r_x$ and 
$r_u$. For example, in level $d=3$ of Figure~\ref{fig:wmmeffic2} and taking
$t=2$, the codes of the nodes $x$ with rank $r=[1,1]$ end with $00$, those 
with ranks $r=[2,3]$ end with $10$, those with ranks $[4,4]$ end with $01$, 
and those with ranks $r=[5,6]$ end with $11$. The differences $r_u-r_x$ are
$+1$ for the termination $00$, $-1$ for $10$, $-2$ for $01$, and $-4$ for 
$11$, the same for all the ranks in the same intervals.

We can then compute the codeword of length $\ell$ in $\Oh{\ell/t}$ chunks of $t$ bits
each, by starting at depth $d=\ell$ and using the formulas to climb by $t$ 
steps at a time until reaching the root (the last chunk may have less than $t$
bits).

For each depth $d$ having $s$ nodes, we store a bitmap $B_d[1..s]$, where
$B_d[r]=1$ if $r$ is the first rank of the interval that ends with the same 
$t$ bits (or the same $d$ bits if $d < t$). A table 
$A_d[rank(B_d,r)]$ then stores those $t$ bits and 
the difference that must be added to each $r_x$ in that interval to make it
$r_u$. Across all the depths, the bitmaps $B_d$ add up to $\Oh{\sigma}$ bits
because $\B$ has $\Oh{\sigma}$ nodes.
Further, there are at most $2^t$ partitions in each depth, so the tables 
$A_d$ add up to $L\cdot 2^t$ entries, each using $\Oh{t+\log\sigma}$ bits:
$t$ bits of the chunk and $1+\log\sigma$ bits to encode $r_u-r_x$, since 
ranks are at most $\sigma$.
In total, we use $\Oh{\sigma + L\,2^t (t+\log\sigma)}$ bits, which setting 
$t = \epsilon L/2$, for any constant $\epsilon>0$, is 
$\Oh{\sigma + 2^{\epsilon L}}$ because $t+\log\sigma = \Oh{L}$ and $L^2 =
\Oh{2^{\epsilon L/2}}$. 
We can then encode any symbol in time
$\Oh{L/t} = \Oh{1/\epsilon}$, that is, a constant.

For decoding we store a table that stores, for every depth $d$ that is a 
multiple of $t$, and every sequence $j$ of $t$ bits, a cell $(d,j)$ with
the value to be added to $r_u$ in order to become $r_x$, where $u$ is any
node at depth $\depth(u)=d$ and $x$ is the node we reach from $u$ if we 
descend using the $t$ bits of $j$.  This table then has $(L/t) \cdot 2^t$ 
entries, each using $\Oh{\lg\sigma}$ bits to encode the value to be added. 
With $t=\epsilon L/2$, the space is $\Oh{2^{\epsilon L}}$ bits and we arrive 
at the desired leaf after $\Oh{1/\epsilon}$ steps (note that our formulas allow 
us identifying leaves). Once we arrive at a leaf at depth $d$, we know the 
codeword length $\ell=d$ and the rank $r=r_x$, so we use the wavelet tree to 
compute the source symbol in constant time.

The obvious problem with this scheme is that it only works if the length $\ell$
of the codeword we find is a multiple of $t$. Otherwise,
in the last step we will try to advance by $t$ bits when the leaf is at less
distance. In this case our computation of $r_x$ will give an incorrect result.

Note from our formulas that the nodes $x$ at depth $d+k$ with
$r_x \le \leaves(d+k)$ are leaves and the others are internal nodes.
Let $u$ be any node at depth $\depth(u)=d$ and $j$ be the bits of a potential
path of length $t$ descending from $u$. If $x$ descends from $u$ by the sequence
$j_k$ of the first $k$ bits of $j$, then the difference 
$g_{d,j}(k) = r_x - r_u$ depends only on $d$, $j$, and $k$ (indeed, our table 
stores precisely $g_{d,j}(t)$ at cell $(d,j)$). Therefore, the nodes $u$ that 
become 
leaves at depth $d+k$ are those with $r_u \le \leaves(d+k) - g_{d,j}(k)$. We can
then descend from node $u$ by a path with $s$ bits $j_s$ iff $r_u > m_{d,j}(s)$,
with
$$ m_{d,j}(s) ~=~ \max_{0 \le k < s} ~ \{ \leaves(d+k) - g_{d,j}(k) \}.$$

We then extend our tables in the following way. For every cell $(d,j)$ we
now store $t$ values $m_{d,j}(s)$, with $s=1,\ldots,t$, and the associated 
values $g_{d,j}(s)$. Note that $m_{d,j}(s) \le m_{d,j}(s+1)$, so this sequence 
is nondecreasing. We make it strictly increasing by removing the smaller $s$ 
values upon ties. To find out how much we can descend from an internal node 
$u$ at depth $d$ by the $t$ bits $j$, 
we find $s$ such that $m_{d,j}(s) < r_u \le m_{d,j}(s+1)$, and then we can 
descend by $s$ steps (and by $t$ steps if $r_u > m_{d,j}(t)$). To descend by
$s$ steps to the descendant node $x$, we compute $r_x = r_u + g_{d,j}(s)$.

We find $s$ with a predecessor search on the $t$ values $m_{d,j}(s)$.
One of the predecessor algorithms surveyed in Section~\ref{sec:basics} runs 
in time $\Oh{\log_w t}$, which is constant in the RAM model with $L = \Oh{w}$
because $t=\epsilon L/2$. Therefore, the encoding time is still 
$\Oh{1/\epsilon}$. The space is now multiplied by $t$ because the
values $m_{d,j}$ and $g_{d,j}$ also fit in $\Oh{\log\sigma}$ bits, and thus
it is still $\Oh{L 2^{\epsilon L/2}} \subseteq \Oh{2^{\epsilon L}}$ bits.

\begin{theorem}
\label{thm:matrices}
Consider an optimal prefix-free code in which all the codewords of length 
$\ell$ come before the prefixes of length $\ell$ of longer codewords in the 
lexicographic order of the reversed binary strings.
We can store such a code in $\Oh{\sigma \log L + 2^{\epsilon L}}$ bits --- possibly after swapping symbols' codewords of the same length --- where $\sigma$ is the alphabet size, $L$ is the maximum codeword length, and $\epsilon>0$ is any positive constant, so that we can encode and decode any codeword in constant time.
The result assumes a $w$-bit RAM computation model with $L=\Oh{w}$.
%Suppose we are given an optimal prefix code in which the codewords' lengths are non-decreasing when they are arranged such that their reverses are in lexicographic order. Let $L$ be the maximum codeword length, so that it is at most a constant times the size of the machine word. Then we can store such a code in $\Oh{\sigma \log L + 2^{\epsilon L}}$ bits --- possibly after swapping symbols' codewords of the same length --- where $\epsilon$ is any positive constant, such that we can encode and decode any symbol in constant time.
\end{theorem}

\section{Experiments} \label{sec:exp}

\newcommand{\EsWiki}{\texttt{EsWiki}}
\newcommand{\EsInv}{\texttt{EsInv}}
\newcommand{\Indo}{\texttt{Indo}}
\newcommand{\ESInvD}{\texttt{ESInvD}}
\newcommand{\HH}{\mathcal{H}}
\newcommand{\LL}{\mathcal{L}}

We have run experiments to compare the solution of Theorem~\ref{thm:matrices0}
(referred to as \wmm\ in the sequel, for Wavelet Matrix Model) with the only
previous encoding, that is,
the one used by Claude et al.~\cite{CNO15} (denoted \tablen). Note that
our codes are not canonical, so other solutions \cite{GNNO15} do not apply.

Claude et al.~\cite{CNO15} use for encoding a single table of $\sigma L$ bits
storing the code of each symbol, and thus they easily encode in constant time.
For decoding, they have tables separated by codeword length $\ell$. In each such
table, they store the codewords of that length and the associated symbol,
sorted by codeword. This requires $\sigma (L+\lg\sigma)$ further bits, and
permits decoding by binary searching the codeword found in the wavelet matrix.
Since there are at most $2^\ell$ codewords of length $\ell$, the binary search
takes time $\Oh{\ell}$.

For the sequence $D$ used in our \wmm, we use binary Huffman-shaped wavelet
trees with plain bitmaps (i.e., not compressed). The structures for supporting 
$rank$/$select$ require $37.5$\% extra space, so the total space is 
$1.37\,\sigma \HH_0(D)$, where $\HH_0(D) \le \lg L$ is the per-symbol 
zero-order entropy of the sequence $D$. We
also add a small index to speed up select queries \cite{NPsea12.1} (at
decoding), which is parameterized with a sampling value that we set to
$\{16, 32, 64, 128\}$. Finally, we store the values $\leaves$ and $\nodes$,
which add an insignificant $L\log\sigma$ bits in total.

We used a prefix of three datasets in \texttt{http://lbd.udc.es/research/ECRPC}.
The first one, \EsWiki, contains a sequence of word
identifiers generated by using the Snowball algorithm to apply stemming to the Spanish Wikipedia.
The second one, \EsInv, contains a concatenation of differentially encoded inverted lists extracted
from a random sample of the Spanish Wikipedia.
The third dataset, \Indo\ was created with the concatenation of the adjacency lists of Web graph
{\tt Indochina-2004}, from {\tt http://law.di.unimi.it/datasets.php}.

Table~\ref{tab:coll0} provides some statistics about the datasets, starting with
the number of symbols in the sequence ($n$) and the alphabet size ($\sigma$).
$\HH(P)$ is the entropy, in bits per symbol, of the frequency distribution $P$
observed in the sequence. This is close to the average length $\ell$ of 
encoded and decoded codewords. The last columns show the maximum codeword 
length $L$ and the zero-order entropy of the sequence $D$, $\HH_0(D)$, in bits 
per symbol. This is a good approximation to the per-symbol size of our 
wavelet tree for $D$.

\begin{table}[t]
\begin{center}
\resizebox{\textwidth}{!}
{\begin{tabular}{|l||r|r|r|r|r|}
\hline
Collection & ~Length & ~Alphabet    & ~Entropy   & ~max code     & Entropy of level \\
           & ~($n$)~& ~size ($\sigma$)~& ~($\HH(P)$) & ~length($L$) & ~entries ($\HH_0(D)$) \\
\hline
\EsWiki	& 200,000,000 & 1,634,145 & 11.12 & 28 & 2.24 \\
\EsInv & 300,000,000 & 1,005,702 & 5.88  & 28 & 2.60 \\
\Indo   & 120,000,000 & 3,715,187 & 16.29 & 27 & 2.51 \\
\hline
\end{tabular}}
\caption{Main statistics of the texts used.}
\label{tab:coll0}
\end{center}
\end{table}

Our test machine has an Intel(R) Core(tm) i7-3820@3.60GHz CPU (4 cores/8 siblings) and 64GB of DDR3 RAM.
It runs Ubuntu Linux 12.04 (Kernel 3.2.0-99-generic). The compiler used was g++ version 4.6.4 and we
set compiler optimization flags to \texttt{-O9}. All our experiments run in a single core and time measures
refer to CPU {\em user-time}. The data to be compressed is streamed from the local disk and also output to disk using the regular buffering mechanism from the OS.

  \begin{figure}[htbp]
  \begin{center}

  \includegraphics[angle=-0,width=0.47\textwidth]{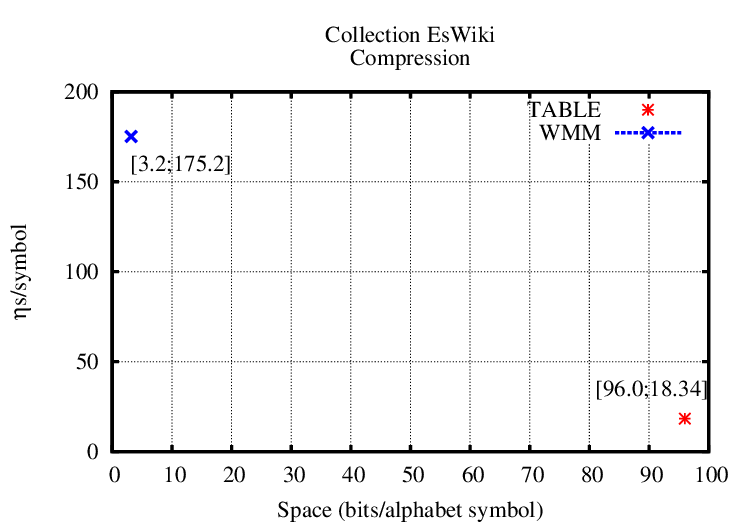}
  \includegraphics[angle=-0,width=0.47\textwidth]{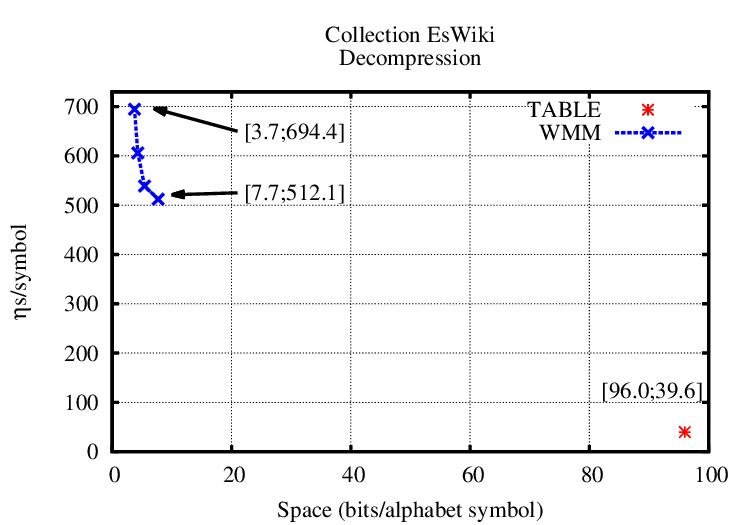}
  \includegraphics[angle=-0,width=0.47\textwidth]{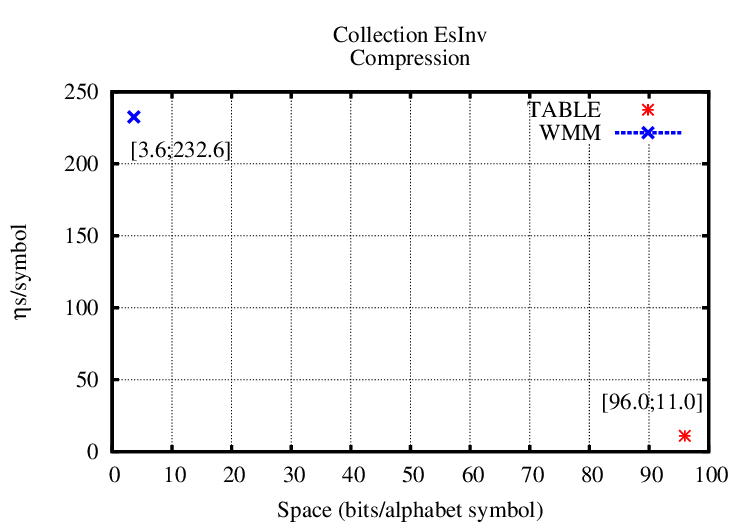}
  \includegraphics[angle=-0,width=0.47\textwidth]{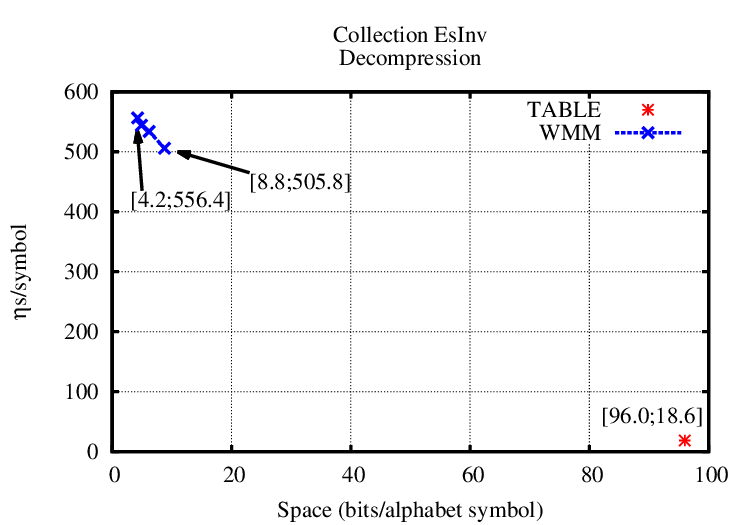}
  \includegraphics[angle=-0,width=0.47\textwidth]{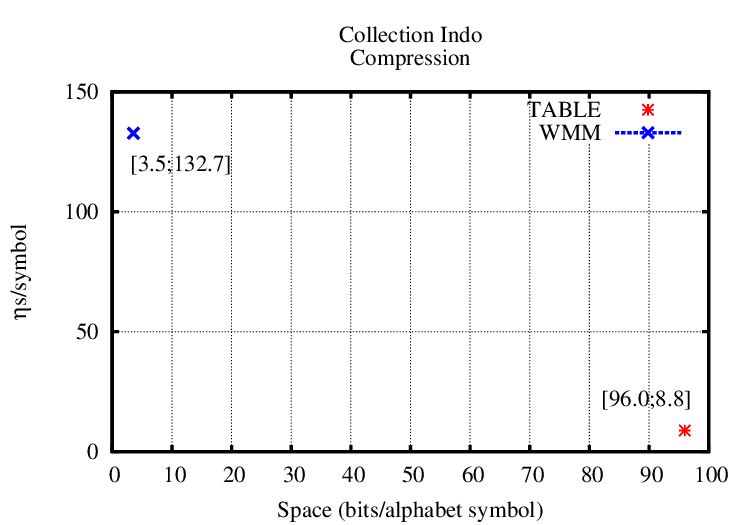}
  \includegraphics[angle=-0,width=0.47\textwidth]{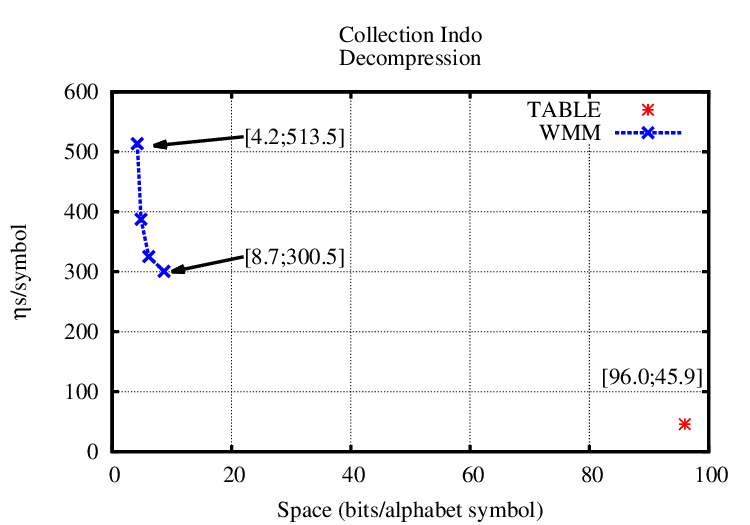}
  \end{center}
  \vspace{-0.3cm}
  \caption{Size of code representations versus compression time (left) and decompression time (right).
  Time is measured in nanoseconds per symbol.}
  \label{fig:exp.enc.dec}

  \end{figure}

Figure~\ref{fig:exp.enc.dec} compares the space required by both
code representations and their compression and decompression times.
As expected, the space per symbol of our new code representation, \wmm, is
close to $1.37\, \HH_0(D)$, whereas that of \tablen\ is close to $2L+\lg\sigma$.
This explains the large difference in space between both representations,
a factor of 23--30 times. For decoding we show the effect of adding the
structure that speeds up select queries.

The price of our representation is the encoding and decoding time. While the
\tablen\ approach encodes using a single table access, in 9--18 nanoseconds,
our representation needs 130--230, which is 10--21 times slower. For
decoding, the binary search performed by \tablen\ takes 20--45 nanoseconds,
whereas our \wmm\ representation requires 500--700 in the slowest and smallest
variant (i.e., 11--30 times slower). Our faster variants require 300--500
nanoseconds, which is still 6.5--27 times slower.

\section{Conclusions}

A classical prefix-free code representation uses $\Oh{\sigma L}$ bits, where
$\sigma$ is the source alphabet size and $L$ the maximum codeword length, and
encodes in constant time and decodes a codeword of length $\ell$ in time
$\Oh{\ell}$.
Canonical prefix codes can be represented in $\Oh{\sigma \log L}$ bits, so that
one can encode and decode in constant time. % under reasonable assumptions.
In this paper we have considered
two families of codes that cannot be put in canonical form. Alphabetic codes
can be represented in $\Oh{\sigma}$ bits, but encoding and decoding
takes time $\Oh{\ell}$. We showed how to store an optimal alphabetic code in $\Oh{\sigma \log L}$ bits such that encoding and decoding any codeword of length $\ell$ takes $\Oh{\min (\ell, \log L)}$ time. % under reasonable assumptions.  
We also showed how to store it in $\Oh{\sigma \log L + 2^{L^\epsilon}}$ bits, 
%under the same assumptions and 
where $\epsilon$ is any positive constant, such that encoding and decoding any such codeword takes $\Oh{\log \ell}$ time.  We thus answered an open problem from the conference version of this paper \cite{SPIRE16}.  We then gave an approximation that worsens the average code
length by a factor of $1+\Oh{1/\sqrt{\log\sigma}}$, but in exchange
requires only $o(\sigma)$ bits and encodes and decodes in constant time. 

We then consider a family of codes where, at any level, the strings leading to 
leaves lexicographically precede the strings leading to internal nodes, if we 
read them upwards.
For those we obtain a representation using $\Oh{\sigma\log L}$ bits and encoding
and decoding in time $\Oh{\ell}$, and even in constant time 
%under reasonable assumptions 
if we use $\Oh{2^{\epsilon L}}$ further bits, where $\epsilon$ is again any positive constant.  We have implemented the simple version of these codes, which
are used for compressing wavelet matrices \cite{CNO15}, and shown that our encodings are
significantly smaller than classical ones in practice (up to 30 times),
albeit also slower (up to 30 times).
We note that in situations when our encodings are small enough to fit in a faster level of the memory hierarchy, they are likely to be also significantly faster than classical ones.

We leave as an open question extending our results to dynamic coding
\cite{Gag04,GKN09,GN09,GN11,GILMN18} and to codes with unequal codeword-symbol
costs \cite{GN09,GL08}. 

\section*{Acknowledgements}

This research was carried out in part at University of A Coru\~na, Spain, while the second author was visiting from the University of Helsinki and the sixth author was a PhD student there. It started at a StringMasters workshop at the Research Center on Information and Communication Technologies (CITIC) of the University of A Coru\~na.  The workshop was funded in part by European Union's Horizon 2020 research and innovation programme under the Marie Sk{\l}odowska-Curie grant agreement No 690941 (project BIRDS).  The authors thank Nieves Brisaboa and Susana Ladra.

The first author was supported by the CITIC research center funded by Xunta de Galicia/FEDER-UE 2014-2020 Program, grant CSI:ED431G 2019/01; by MICIU/FEDER-UE, grant BIZDEVOPSGLOBAL: RTI2018-098309-B-C32; and by Xunta de Galicia/FEDER-UE, ConectaPeme grant GEMA: IN852A 2018/14. 
The second author was supported by Academy of Finland grants 268324 and 250345 (CoECGR), Fondecyt Grant 1-171058, and NSERC grant RGPIN-07185-2020. The fourth author was supported by PRIN grant 2017WR7SHH, and by the INdAM-GNCS Project 2020 {\sl MFAIS-IoT}. The fifth author was supported by Fondecyt Grant 1-200038, Basal Funds FB0001, and ANID -- Millennium Science Initiative Program -- Code
ICN17\_002, Chile. 

%\section*{Declarations of Interest: None}

\end{document}